\documentclass[aps,prl,twocolumn,groupedaddress]{revtex4}

\usepackage{graphicx}
\usepackage{dcolumn}
\usepackage{bm}

\begin{document}

\title{\large  Magnetoresistance and gating effects in ultrathin NbN-$\rm Bi_2Se_3$ bilayers.}

\author{Gad Koren}
\email{gkoren@physics.technion.ac.il} \affiliation{Physics
Department, Technion - Israel Institute of Technology Haifa,
32000, ISRAEL} \homepage{http://physics.technion.ac.il/~gkoren}

\date{\today}
\def\bfig {\begin{figure}[tbhp] \centering}
\def\efig {\end{figure}}

\normalsize \baselineskip=8mm  \vspace{15mm}


\begin{abstract}

\noindent\textbf{Abstract}\\
Ultrathin $\rm Bi_2Se_3$-NbN bilayers comprise a simple proximity system of a topological insulator and an s-wave superconductor for studying gating effects on topological superconductors. Here we report on 3 nm thick NbN layers of weakly connected superconducting islands, overlayed with 10 nm thick $\rm Bi_2Se_3$ film which facilitates enhanced proximity coupling between them. Resistance versus temperature of the most resistive bilayers shows insulating behavior but with signs of superconductivity. We measured the magnetoresistance (MR) of these bilayers versus temperature with and without a magnetic field H normal to the wafer (MR=[R(H)-R(0)]/\{[R(H)+R(0)]/2\}), and under three electric gate-fields of 0 and $\pm2$ MV/cm.
The MR results showed a complex set of gate sensitive peaks which extended up to about 30 K.
The results are discussed in terms of vortex physics, and the origin of the different MR peaks is identified and attributed to flux-flow MR in the isolated NbN islands and the different proximity regions in the $\rm Bi_2Se_3$ cap-layer. The dominant MR peak was found to be consistent with enhanced proximity induced superconductivity in the topological edge currents regions. The high temperature MR data suggest a possible pseudogap phase or a highly extended fluctuation regime.\\


\end{abstract}

\maketitle

\section{1. Introduction}
\normalsize \baselineskip=6mm  \vspace{6mm}

Surface edge states of topological superconductors (TOS) are expected to support zero energy modes or Majorana fermions which are robust against disorder and decoherence \cite{KaneRMP,FuKane}. These zero energy excitations could therefore be useful in potential application in spintronics and quantum computing \cite{Pesin,Kitaev}. Bulk TOS such as copper doped $Bi_2Se_3$ should have been the simplest materials to study TOS properties, but complications due to their inherent inhomogeneity \cite{Kriener} and the presence of possible superconducting impurity phases such as $CuSe_2$ \cite{AndoRev}, make them less attractive for such investigations. An alternative way for realizing TOS is by inducing superconductivity in a topological insulator or in semiconductor-nanowires with strong spin-orbit interaction via the proximity effect (PE) \cite{Koren1,LiLu,Kouwenhoven,Heibloom}. Unconventional superconductivity in these systems, such as revealed by the presence of zero bias conductance peaks (ZBCP), indicates zero energy bound states that might be due to Majorana zero energy modes, but could also originate in zero energy Andreev bound states. It is hard to distinguish between these two different phenomena, and efforts are ongoing in order to achieve this goal \cite{Xu,Berg}. \\

Spatial sharpness of the boundary region between the superconductor and the topological or semi-conducting material is also essential in order to distinguish between the near-zero-energy end states originating in Andreev bound states, and the Majorana zero energy modes in the topological case \cite{Brouwer}. Since this boundary is generally created by gating in the experiments, its unavoidable gradual spatial change adds more uncertainty to the interpretation of the observed ZBCPs as due to the Majorana modes \cite{Kouwenhoven,Heibloom,Marcus}. The role of gating in these nanowire-superconductor experiments was further investigated and a variety of additional phenomena such as ZBCP oscillations versus gate voltage and magnetic field were observed and interpreted in the context of Majorana modes as well as alternatives such as Kondo and disorder effects  \cite{Churchill}. Hence, gating is sufficiently important in these studies and we decided to use a simple proximity system of a bilayer comprising of a topological insulator and a 2D superconductor for studying gating effects on topological superconductors \cite{KorenSUST}. In continuation to our previous studies of $Bi_2Se_3-NbN$ junctions \cite{Koren2,Koren3} we used ultra-thin bilayers of this system, even thinner than used before \cite{KorenSUST}, where the superconducting $NbN$ islands with weak-links in between them are overlayed with a thicker $Bi_2Se_3$ layer which enhances the coupling between the $NbN$ islands via the (inverse) proximity effect. Here we report on gating effects on the magnetoresistance (MR) of this hybrid system, which shows a highly gate sensitive, non-monotonous behavior versus temperature.\\

\begin{figure} \hspace{-20mm}
\includegraphics[height=4cm,width=8cm]{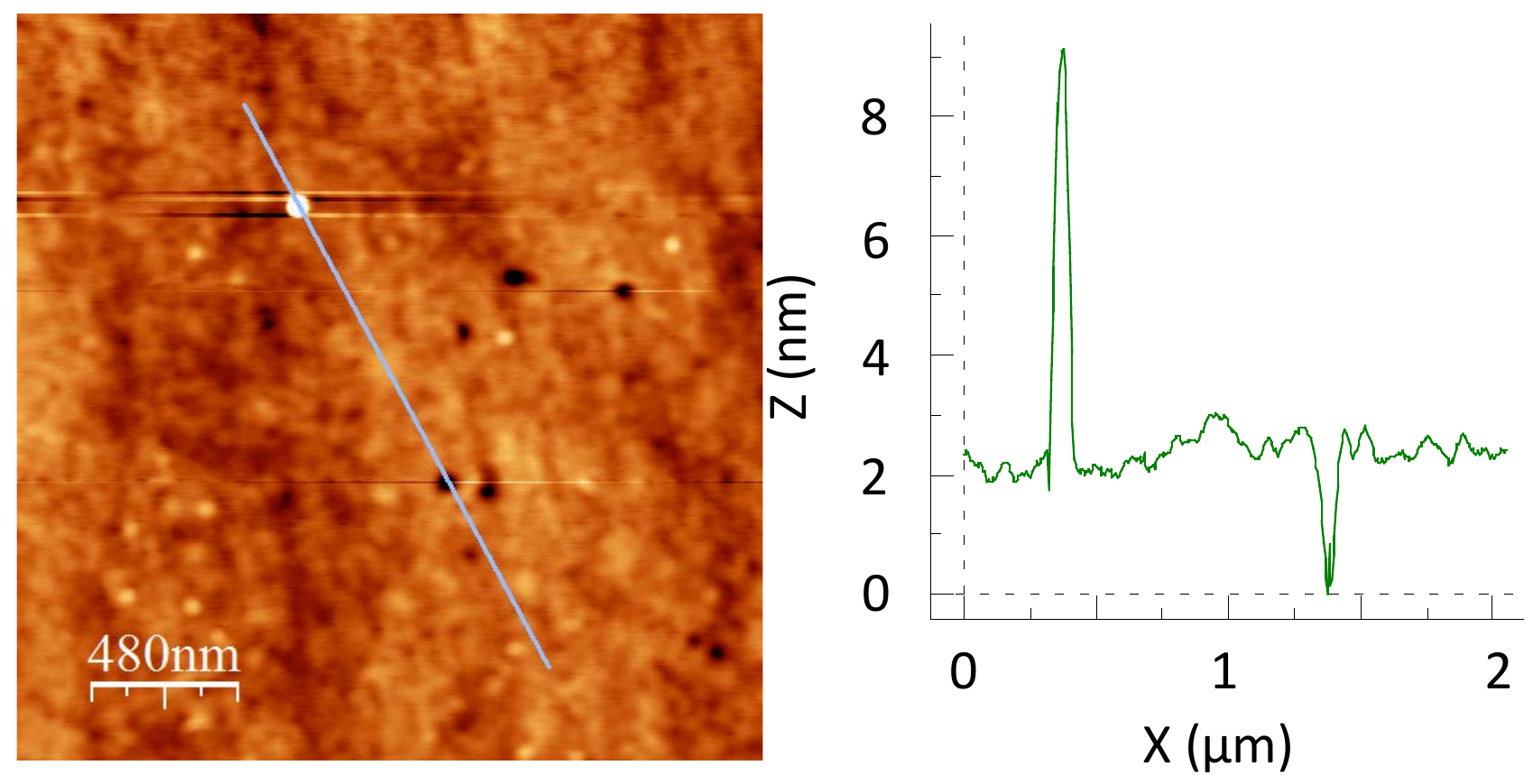}
\hspace{-20mm} \caption{\label{fig:epsart} (Color online) Atomic force microscope (AFM) image of nominally 3 nm thick NbN layer, together with a height profile along the line shown in the image. The two horizontal lines near the 9-2=7 nm tall particulate are artifacts due to the AFM inability to follow smooth traces near a big disturbance (in tapping-mode here while scanning along the x-direction).}
\end{figure}

\begin{figure} \hspace{-20mm}
\includegraphics[height=5cm,width=8cm]{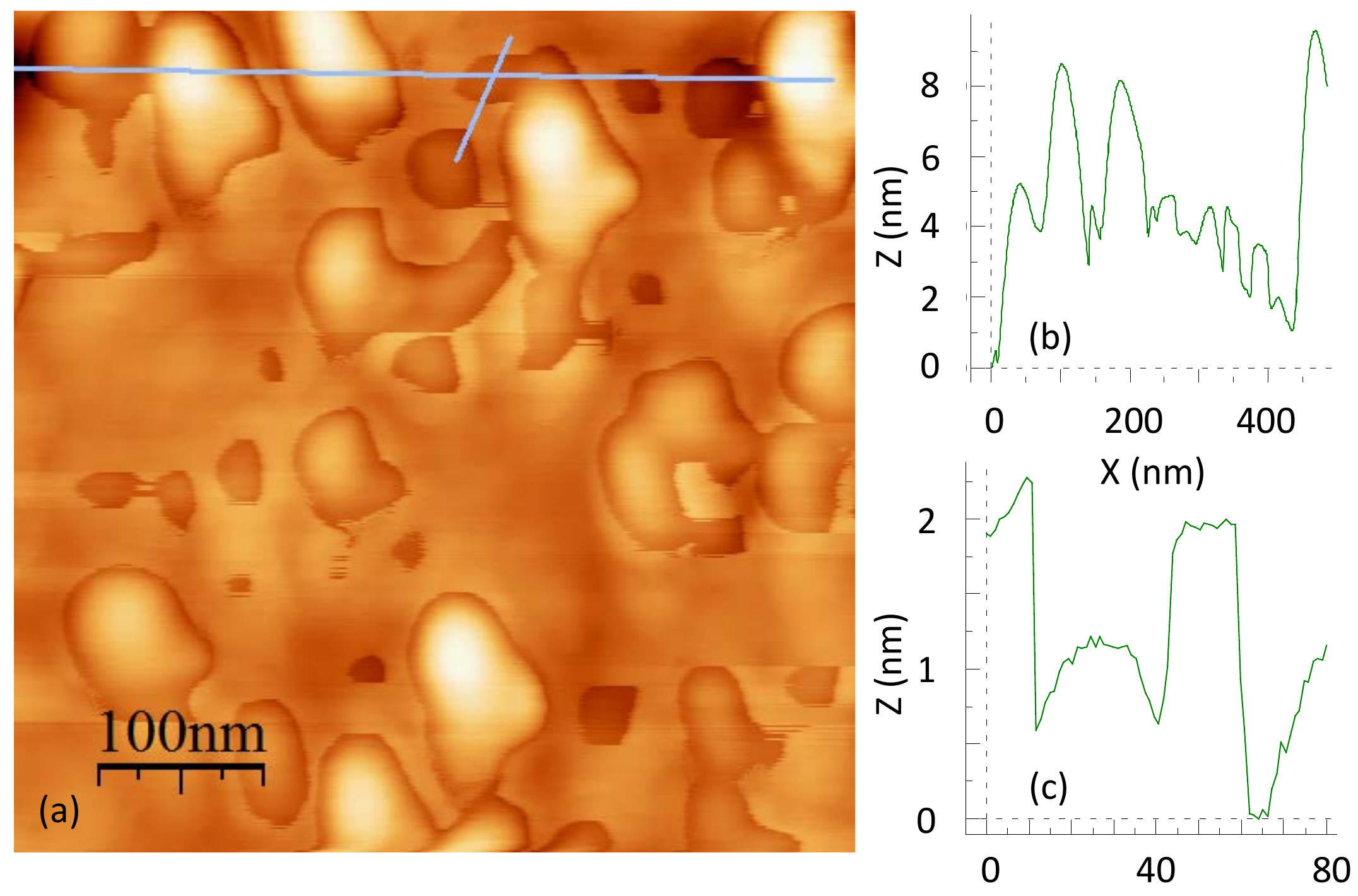}
\hspace{-20mm} \caption{\label{fig:epsart} (Color online) AFM image of a 10 nm  $Bi_2Se_3$ on a 3 nm NbN bilayer (a), together with two height profiles (b) and (c) along the long and short lines shown in the image, respectively.}
\end{figure}

\section{2. Experimental methods}
\subsection{2.1. Preparation and surface morphology of the films and bilayers }
\normalsize \baselineskip=6mm  \vspace{6mm}

The NbN and $Bi_2Se_3$ thin films were prepared as described in detail previously \cite{KorenSUST}. Briefly, laser ablation deposition was used where the NbN films were deposited under 30 mTorr of $\rm N_2$ gas flow and at 600 $^0$C heater block temperature, while the $Bi_2Se_3$ layers were deposited under vacuum and at 300 $^0$C. All films and bilayers were deposited on fused silica wafers and yielded grainy films with rms roughness of $\sim$10\% of the films thickness. Fig. 1 shows an atomic force microscope (AFM) image of the surface morphology of a 3 nm thick NbN film, together with a typical height profile along the line shown in the image. While the lateral size of the NbN grains is $\sim$50-100 nm, one can see larger bright areas of at least 1000 nm in size which represent thicker superconducting NbN islands. These are separated by broad darker areas which constitute thinner weak-links between these islands. This islands structure that comprises a network of strong superconducting regions connected by weak-links is essential for the present study, otherwise a superconducting short will mask all our transport data.  SEM images of similar NbN films on glass showing their grainy nature can also be seen in \cite{Xin-kang}. In addition, Fig. 1 shows a few 1-3 nm deep holes of about 60 nm diameter while the overall rms roughness of this film is $\sim$0.35 nm.  \\

All the $Bi_2Se_3-NbN$ bilayers in the present study were obtained using an \textit{"in-situ"} process where both the 3 nm thick $NbN$ film and the 10 nm thick $Bi_2Se_3$ cap layer were prepared in the same deposition run without breaking the vacuum. This kept the interface between the NbN and $Bi_2Se_3$ layers protected against contamination and oxidation which occur if the NbN surface is exposed to air \cite{KorenSUST,Darlinski}. Fig. 2 shows an AFM image of the surface morphology of such a bilayer (a), together with two height profiles (b) and (c) along the long and short lines shown in (a), respectively. The bilayer shows taller grains on a smoother background, with grains height of 4-8 nm and some holes, the deepest of which is of about 5 nm deep. Shallower holes are also visible which reveal a step-like depth profile as seen in (c), where the step heights are of about 1 or 2 nm in agreement with one or two quintuples of the $Bi_2Se_3$ structure. The overall rms roughness of this bilayer is $\sim$1.1 nm. \\

\subsection{2.2.  The gates and contacts geometry }
\normalsize \baselineskip=6mm  \vspace{6mm}

Transport measurements were done using an array of 40 gold coated spring loaded spherical tips for the 4-probe measurements on 10 different locations on the wafer (Ci with i=1 to 10, 4 contacts for each location). Fig. 3 depicts a schematic drawing of the bilayer, gates positions and 20 representative contacts. Contacts C1 and C10 are separated from the bilayer by two scratches, while the 4 contacts of each are shorted and connected to the top and bottom gates, respectively. Since the fused silica substrate is 1.5 mm thick and its dielectric constant is low, the bottom gate is useless here and is shorted to the bilayer and ground, as can be seen in the schematic cross-section of the bilayer and gates in Fig. 4. The insulation layer of the top gate is made of a much thinner PMMA resist of 1.3 $\mu$m thickness. This allows for an electric field of $\pm$2 MV/cm to be obtained when a voltage of $\mp$100 V  is applied to the top gate (opposite polarity). This will be marked in the following by $\mp$100 Vg. Fig. 4 shows two weakly connected NbN islands (red dots) together with their depletion layer (blue dots on white background) in the $Bi_2Se_3$ layer near the interface with the NbN islands. Proximity induced superconductivity also occurs in this layer and can extend up to the top surface of the  $Bi_2Se_3$ film \cite{KorenSUST}. \\

\begin{figure} \hspace{-20mm}
\includegraphics[height=7cm,width=8cm]{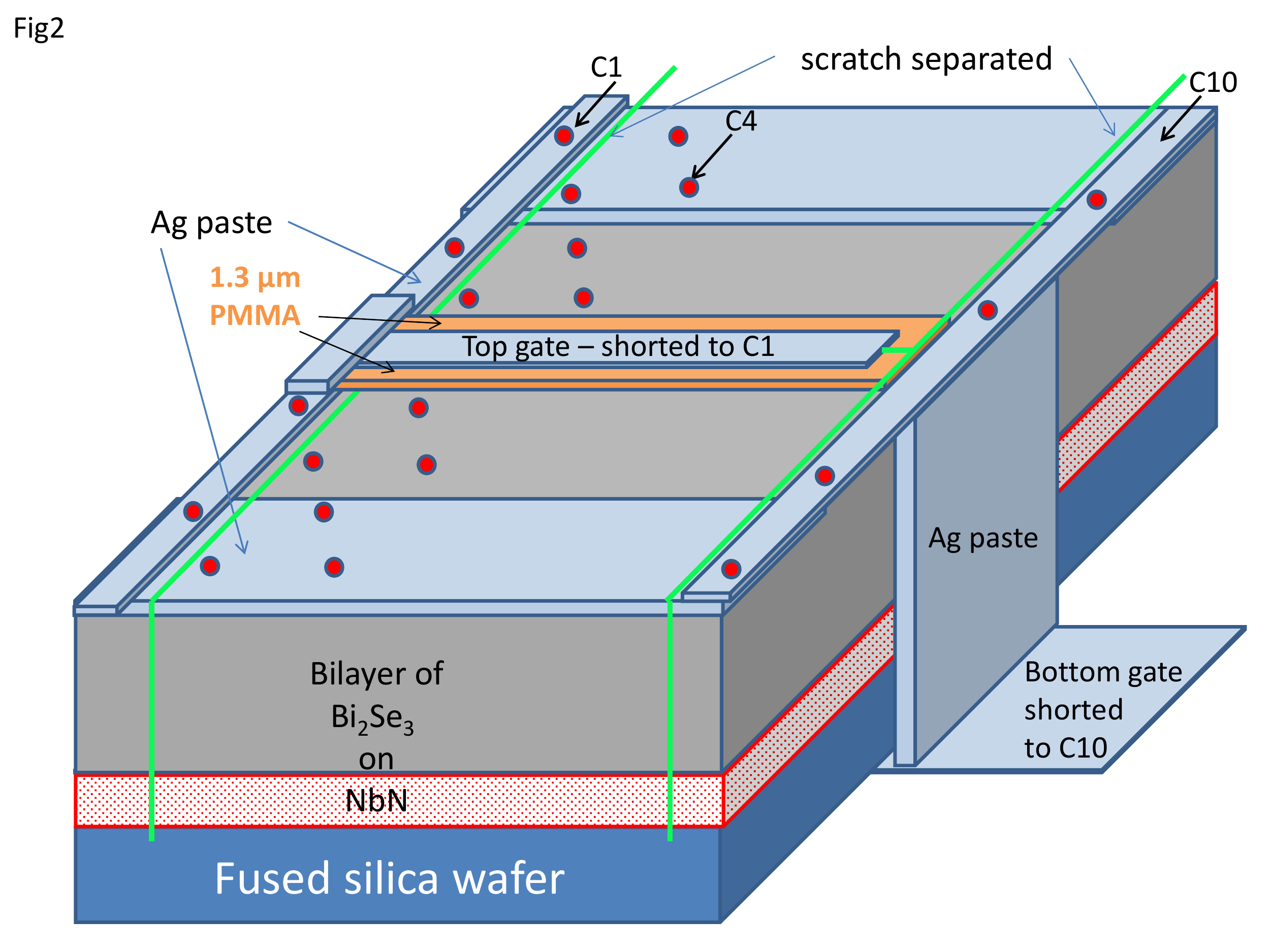}
\hspace{-20mm} \caption{\label{fig:epsart} (Color online) (a) A schematic drawing of the $Bi_2Se_3$ on NbN bilayer, together with the top and bottom gates and some of the C\textit{i} contacts where \textit{i}=1 to 10 (each C\textit{i} has 4-point contacts). The C1 and C10 contacts are separated from the bilayer and dedicated for contacting the top and bottom gates, respectively. }
\end{figure}

\begin{figure} \hspace{-20mm}
\includegraphics[height=6cm,width=8cm]{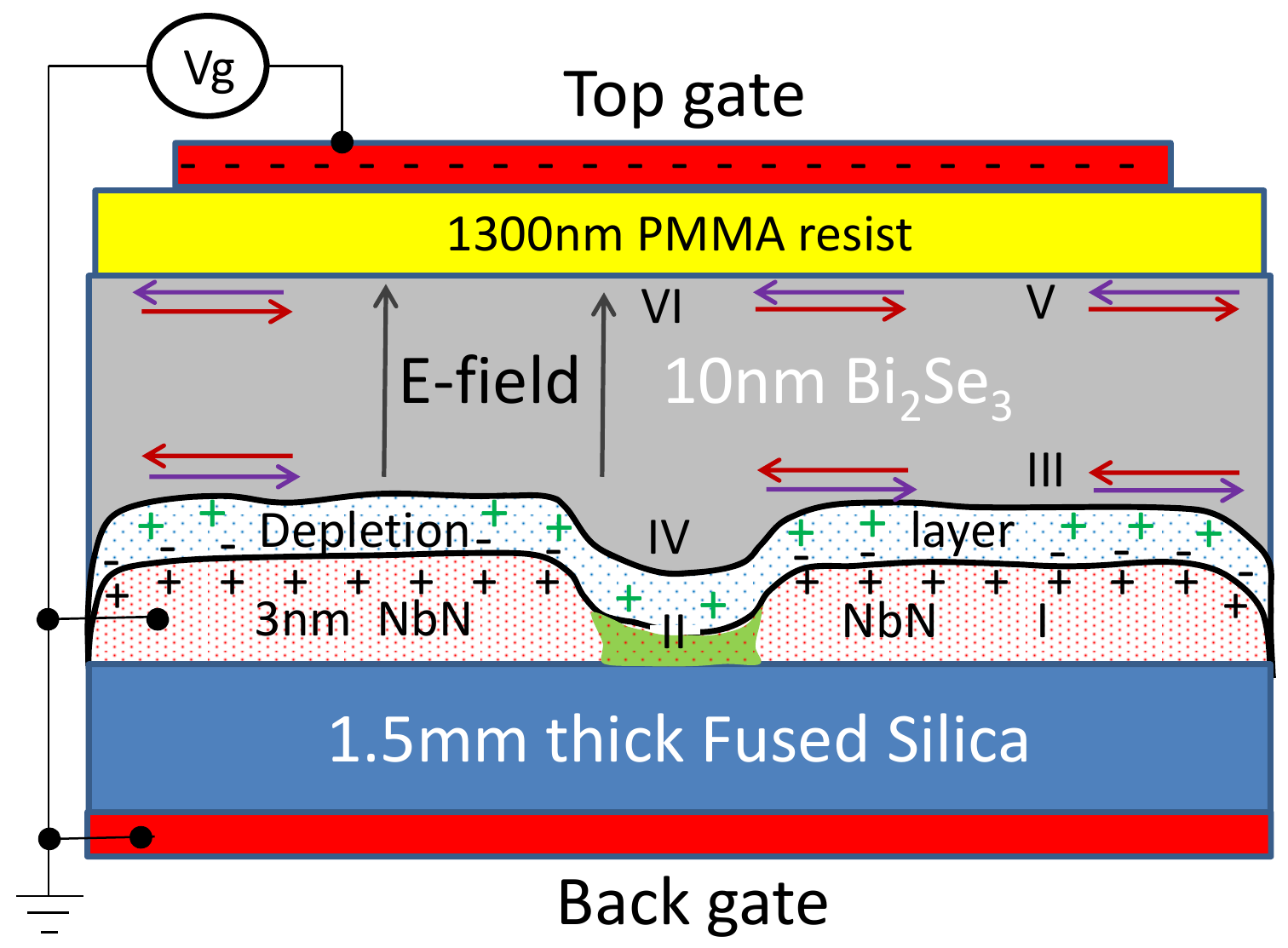}
\hspace{-20mm} \caption{\label{fig:epsart} (Color online) A schematic cross-section of the bilayer and gates. Two superconducting NbN islands marked by red dots are weakly connected by the non-superconducting green shaded area. Helical edge currents for the two spin orientations of the $Bi_2Se_3$ layer are depicted by the horizontal arrows on the top and bottom surfaces. The depletion layer in the $Bi_2Se_3$ is marked by the blue-dots. The back gate is grounded and shorted to the bilayer in the present study, and a negative top gate voltage -Vg corresponds to a positive electric field E.}
\end{figure}

\section{3. Results and discussion}

\subsection{3.1. Comparing bilayers to NbN films}

We start this study by repeating the main experiment of our previous work \cite{KorenSUST} of comparing the properties of a $Bi_2Se_3$-NbN  bilayer to those of a reference NbN film. This time however, half as thin bilayer and reference film are used, in order to enhance the weak-link behavior of both and minimize superconducting shorts. No gates were prepared on the wafer used for this experiment in order to minimized its exposure to ambient air which is detrimental to the NbN film \cite{Darlinski}. A bilayer of 10 nm $Bi_2Se_3$ on 3 nm NbN was deposited on half the fused silica (FS) wafer, while the 3 nm thick NbN reference film was deposited on the other half. The two halves were separated by a scratch as shown in Fig. 2 (a) of Ref. \cite{KorenSUST}. Scratch separation rather than photolithography was used in order to avoid deterioration of the films. The transport results on this wafer are presented in Figs. 5 and 6. As observed previously, the R versus T results show that the onset $T_c$ of the reference NbN film is higher by about 1 K than that of the bilayer. This indicates a standard proximity effect (PE) where the normal electrons of the $Bi_2Se_3$ suppress superconductivity in the NbN layer of the bilayer \cite{KorenSUST}. The inverse PE however, where enhanced superconductivity in the bilayer occurs due to pair penetration into the $Bi_2Se_3$, is absent now, at least down to the minimum temperature used in this study (1.8 K). This, together with the fact that both bilayer and reference film do not reach zero resistance down to 1.8 K, is a good indication that the weak-links in both are actually weaker, as originally planned. The inset to Fig. 5 shows the magnetoresistance of the bilayer and reference film versus temperature. It shows that the $T_c$ onset (determined by the rise of the MR) of the reference film is again, larger than that of the bilayer. Fig. 6 exhibits a zoom-in on the low MR data and its extension to higher temperatures. It shows that the MR onset of the reference NbN layer (C7 RL) is at about 5 K, in agreement with the R versus T data of Fig. 5. It also shows that above 5 K, its MR is zero to within the noise of the measurement. This is different from the bilayer data of Fig. 6 (C2 and C4), where a non zero, positive MR signal of a few percents is observed up to 50 K. For comparison we show an MR signal of a 10 nm thick $Bi_2Se_3$ film on FS, which is very similar to the bilayer data. We thus conclude that above 5 K in this bilayer, the MR signal originates in the $Bi_2Se_3$ cap layer with no noticeable effect of the underlying NbN layer.\\

\begin{figure} \hspace{-20mm}
\includegraphics[height=6cm,width=8cm]{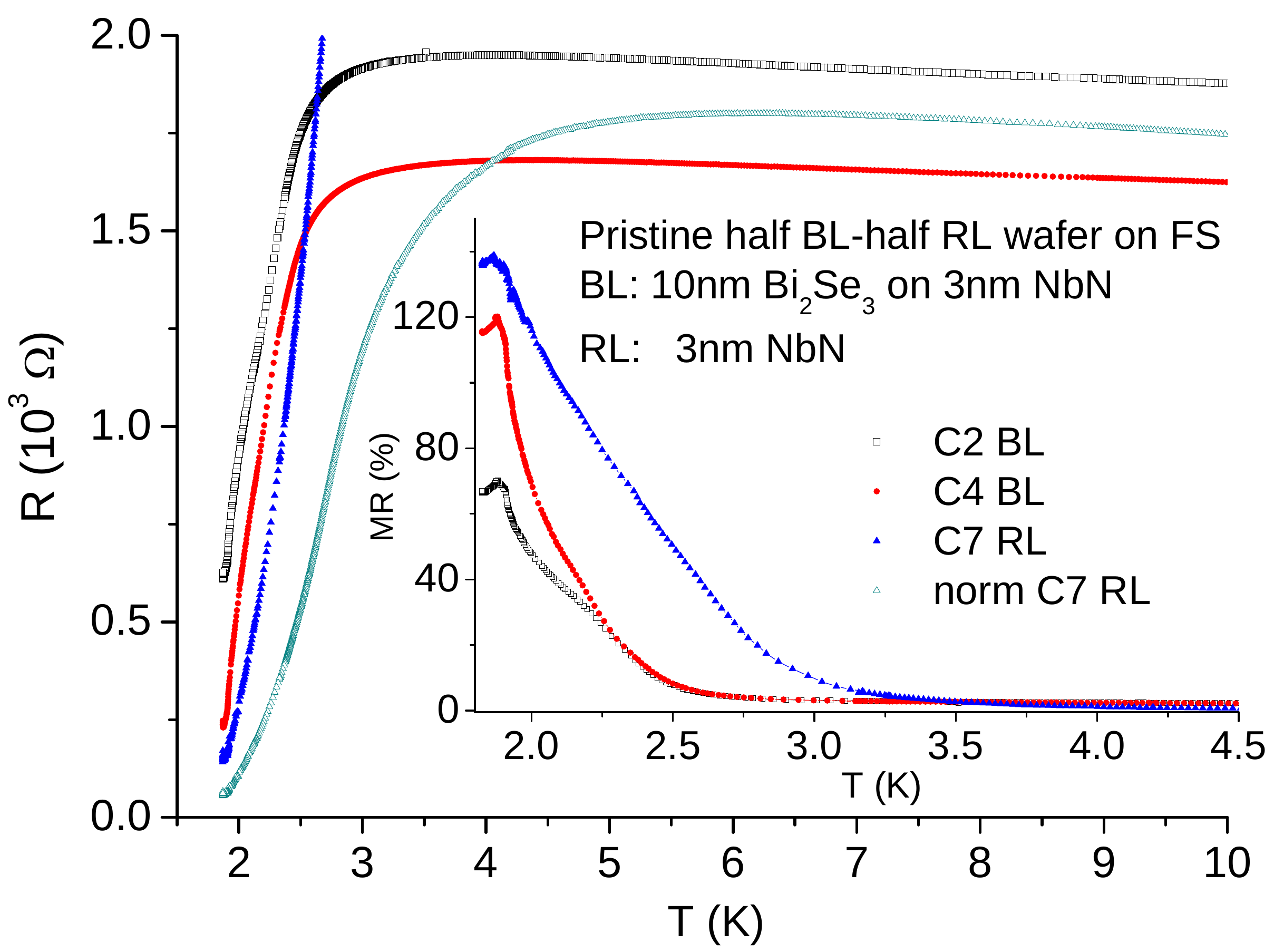}
\hspace{-20mm} \caption{\label{fig:epsart} (Color online) Resistance versus temperature of a pristine, half bilayer (BL) - half reference layer (RL) on a fused silica wafer under 0 Vg and 0 T. The BL is a 10 nm $Bi_2Se_3$ on 3 nm NbN bilayer and the RL is a 3 nm NbN film. Also shown is RL data normalized to the BL data at 10 K (norm C7 RL). Inset: The magnetoresistance at 0 and 1 T magnetic field normal to the wafer versus temperature.   }
\end{figure}

\begin{figure} \hspace{-20mm}
\includegraphics[height=6cm,width=8cm]{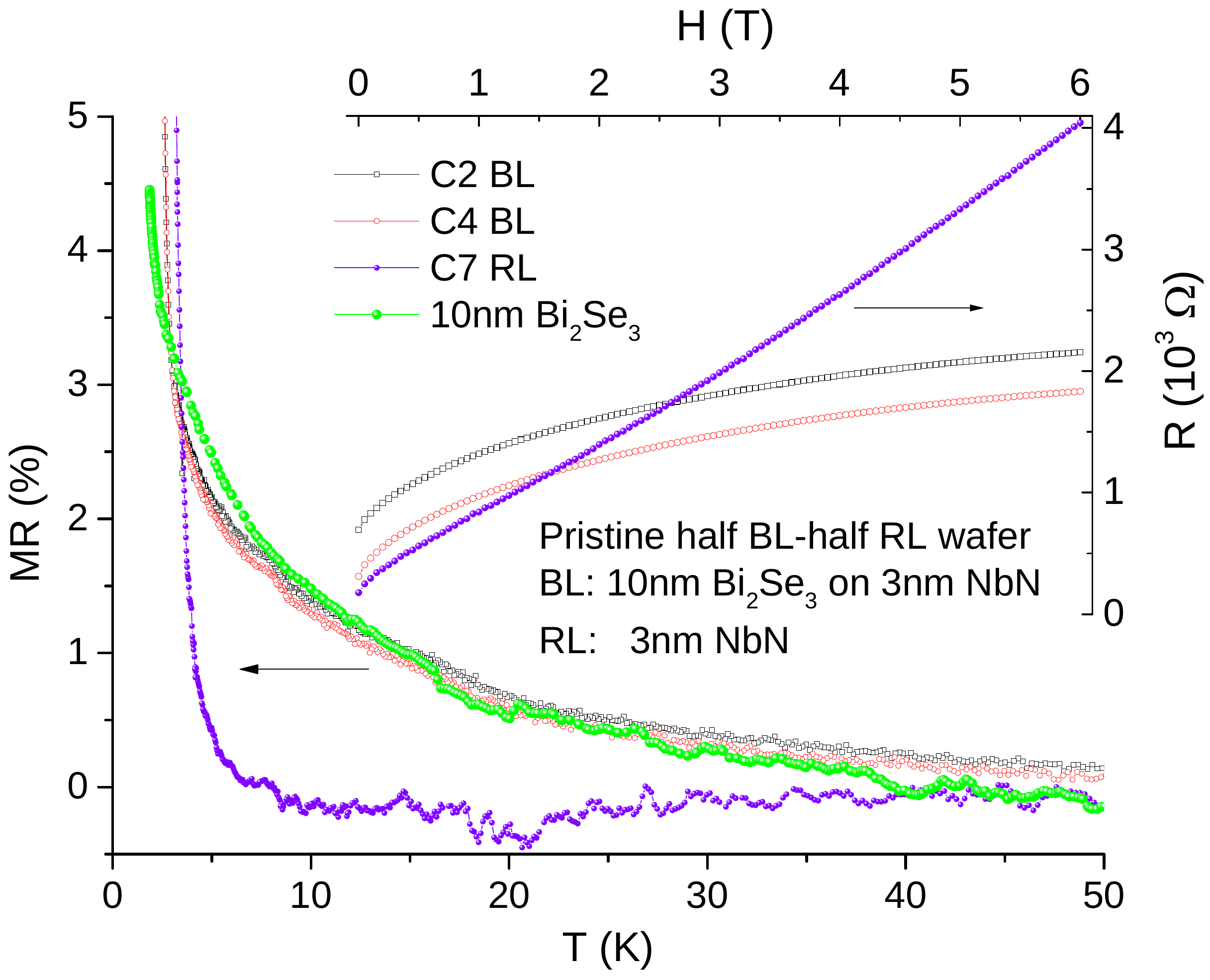}
\hspace{-20mm} \caption{\label{fig:epsart} (Color online) Main panel: Zoom in on the high temperature tail of the MR data of the bilayer and reference film of the inset to Fig. 5. Also added for comparison is the MR versus T of a bare 10 nm thick film of $Bi_2Se_3$ on fused silica. Inset: Resistance versus magnetic field at 1.88 K of the same BL and RL contacts.   }
\end{figure}

The inset to Fig. 6 shows the resistance versus magnetic field H normal to the wafer at 1.88 K, for the same bilayer and reference film contacts. All the resistances are increasing with field, but while the bilayer resistance tends to saturate at high fields, that of the reference film increases quite linearly with H. The later indicates flux-flow resistance in a superconductor as the mechanism for the linearly increasing resistance. If we go back now to the inset of Fig. 5, we see that this effect is stronger in the reference NbN film than in the bilayer. The peak MR at $\sim$1.9 K in all contacts is due to maximum flux flow which results from compensation between increased vortex generation and pinning effect with decreasing temperature. Another interesting result in the inset to Fig. 5, is that in the C2 BL contact there is a clear knee in the MR data at T$\sim$2.1 K which indicates two different MR behaviors below and above it. We attribute the MR above this temperature to the stronger superconducting NbN islands as seen in Fig. 1, while below it, where the weak-links (the darker areas in Fig. 1) also become superconducting due to proximity coupling via the $Bi_2Se_3$, the MR increases further on lowering the temperature. In the following we shall see more supporting evidence for this interpretation. \\

\begin{figure} \hspace{-20mm}
\includegraphics[height=6cm,width=8cm]{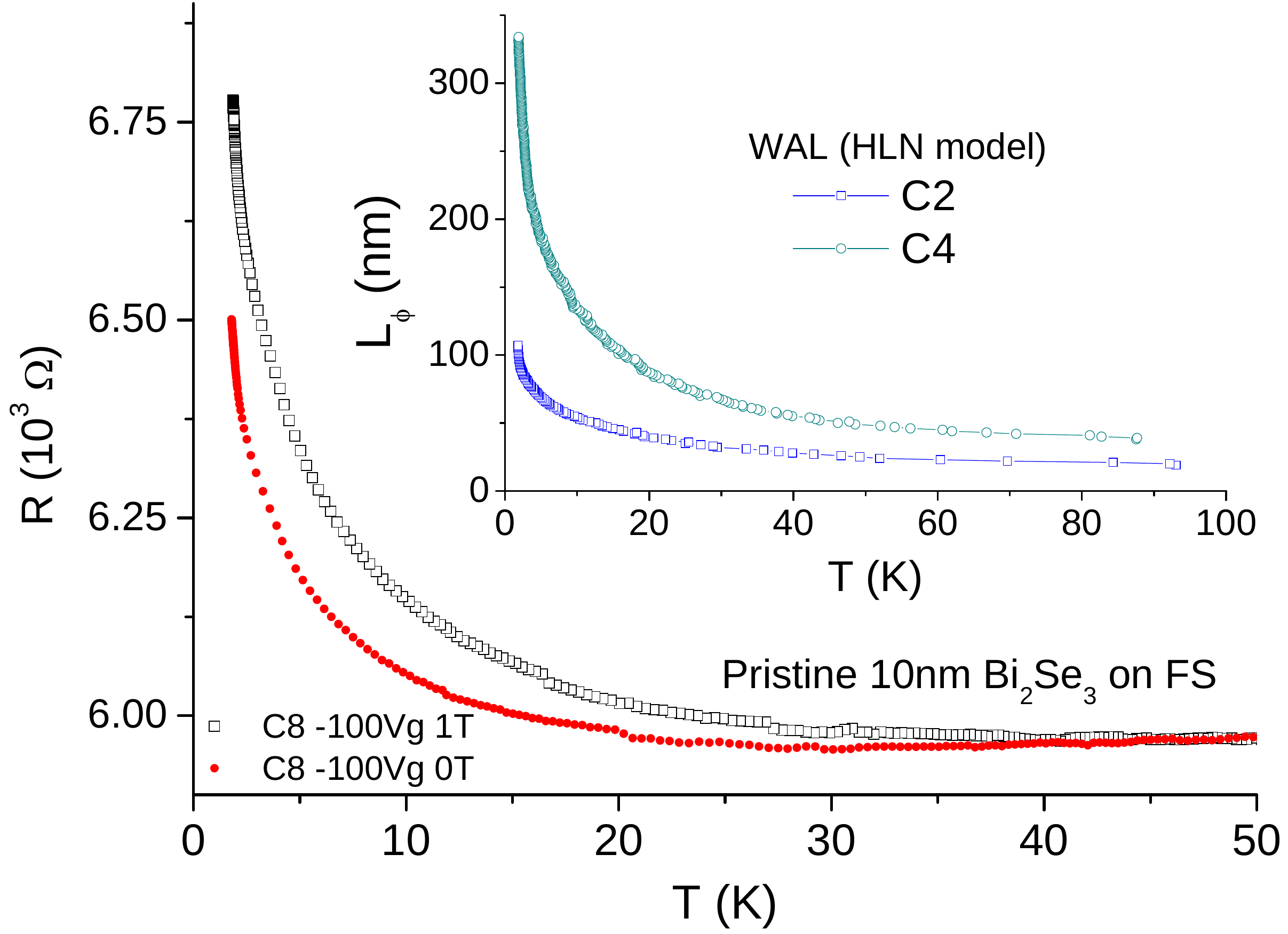}
\hspace{-20mm} \caption{\label{fig:epsart} (Color online) Resistance versus temperature of a 10 nm thick $Bi_2Se_3$ reference film on FS with and without a magnetic field of 1T. The inset shows the phase coherence length $L_\phi$ calculated for another 10 nm $Bi_2Se_3$ film assuming weak anti-localization (WAL) and using the HLN model \cite{HLN}.  }
\end{figure}

\subsection{3.2. R and MR of reference $Bi_2Se_3$ films }
\normalsize \baselineskip=6mm  \vspace{6mm}

For understanding the MR of the bilayers one needs in addition to the MR of the 3 nm NbN reference film, also the MR behavior of a second reference film of 10 nm $Bi_2Se_3$. Such films were prepared and Fig. 7 presents resistance versus temperature results of a 10 nm $Bi_2Se_3$ film under 0 and 1 T magnetic field and -100 V gate voltage. The resistivity of this film is $\sim$6 m$\Omega$cm or 6k$\Omega$ per square, which corresponds to an electron density of about $10^{17}$ cm$^{-3}$ \cite{Butch}. Thus the electron doping of this film is quite low, which results from the Se rich target used in its deposition process \cite{KorenSUST}. The R versus T data of Fig. 7 was used in order to calculate the magnetoresistance of this film and the result has already been presented in the main panel of Fig. 6. We found that MR versus T curves measured on the $Bi_2Se_3$ film were insensitive to the gate voltage used here  ($\mp$100 Vg or $\pm$2 MV/cm) to within the noise of the measurements. Steinberg \textit{et al.} \cite{Steinberg} did observe significant gating effects in a 20 nm thick $Bi_2Se_3$ film, but their maximum electric field was about 40 times higher than what we used here. Thus, the presently used maximum gate voltage  was insufficient to have any significant effect on the MR of the bare 10 nm $Bi_2Se_3$ film. Fig. 8 shows the resistance versus magnetic field of this film at 1.85 K. The inset shows the data obtained on the pristine film, while the main panel shows the data of a one week old film which was kept under dry air. Clear aging effect was observed when the measurements were repeated after one week, as can be seen from the increased resistance of about 25\%. Besides this, both sets of measurements show a linear R versus H behavior at low fields up to about 0.1 T  with saturation at higher fields. \\

\begin{figure} \hspace{-20mm}
\includegraphics[height=7cm,width=8cm]{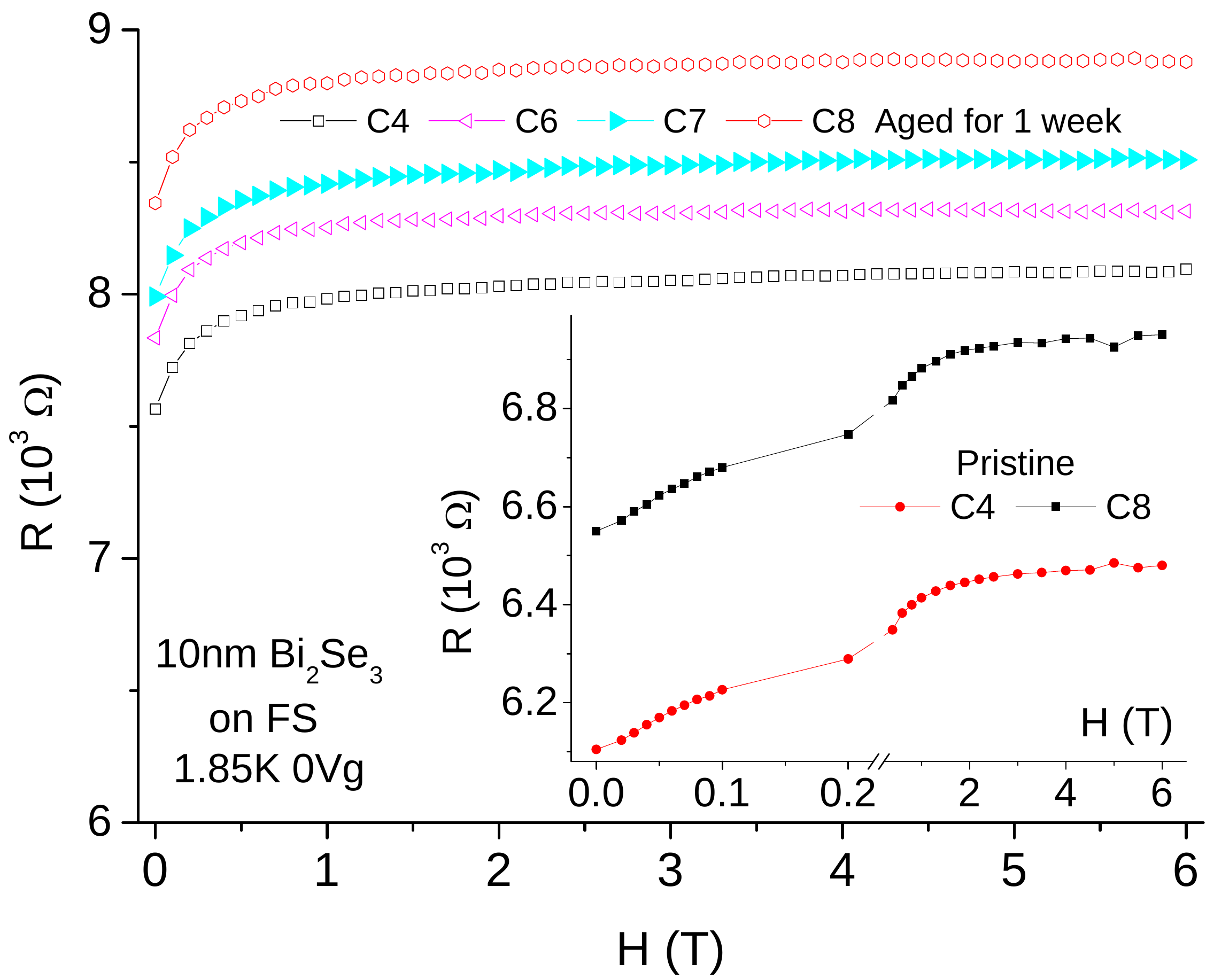}
\hspace{-20mm} \caption{\label{fig:epsart} (Color online) Resistance versus magnetic field at 1.85 K for the 10 nm $Bi_2Se_3$ film. The inset shows results measured on the pristine film while the main panel shows data obtained after the film was aged in dry air ambient for one week. Clearly, the resistance of the aged film went up due to deterioration of its surface layer.}
\end{figure}

The linear MR in $Bi_2Se_3$ is generally attributed to weak anti-localization (WAL) \cite{Bergmann,Steinberg}. If we assume that this is correct, we can use the HLN model \cite{HLN} to calculate the phase coherence length $L_\phi$ as a function of temperature from our data. For a field of H = 1 T and $L_\phi$ in nm, the HLN model yields:

\begin{equation}
1.2\times 10^5 \frac{R(H)-R(0)}{R(0)R(H)} \cong \psi(\frac{1}{2} + \frac{156}{L_\phi^2}) -
ln(\frac{156}{L_\phi^2})
\end{equation}
where $\psi$ is the digamma function. Since both $R(H=1 T)$ and $R(0)$ were measured as a function of temperature, Eq. (1) allows us to extract $L_\phi$ which is thus also a function of temperature. The result is shown in the inset to Fig. 7 for another 10 nm thick $Bi_2Se_3$ film on FS. One can see that the resulting $L_\phi$ is very sensitive to the contact location on the wafer, probably because the coherent WAL scattering process is very sensitive to even slight inhomogeneities in the film. Also, since C2 is closer to the edge of the wafer than C4, edge effects can affect the resulting $L_\phi$. Both however, decrease very rapidly versus T at low T and then decay more gradually. Again, this is not a proof that WAL occurs here, but if it does, these are the calculated $L_\phi$(T) curves from our measurements.\\

\subsection{3.3. Reference bilayer of Au on NbN }
\normalsize \baselineskip=6mm  \vspace{6mm}

\begin{figure} \hspace{-20mm}
\includegraphics[height=6cm,width=8cm]{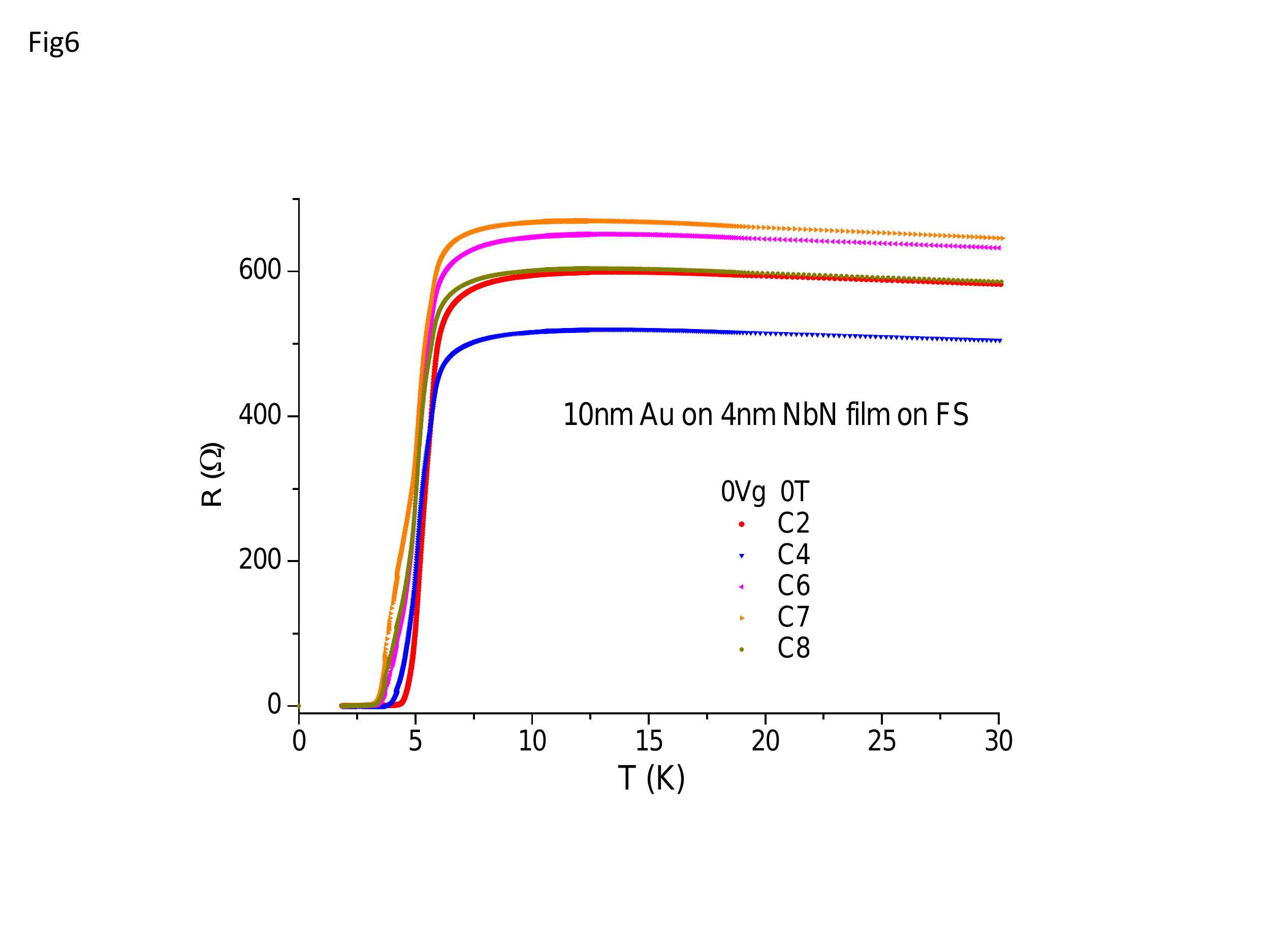}
\hspace{-20mm} \caption{\label{fig:epsart} (Color online) Resistance versus temperature of a reference bilayer of 10 nm Au on 4 nm NbN under zero magnetic field and 0 Vg.  }
\end{figure}

In this section we study the temperature dependence of R and MR of a 10 nm Au on 4 nm NbN bilayer on FS. This provides information on the proximity effect of NbN with a good metal, and allow for a comparison with the $Bi_2Se_3$-NbN bilayers which will be presented and discussed in detail in the next section. We chose to have a bit thicker NbN layer, of 4 nm instead of 3 nm, in this reference bilayer with 10 nm gold cap layer, since we worried that the ball-like gold grains will not fully protect the NbN layer from oxidation. The transport results of R and MR under different magnetic fields and gate voltages are shown in Figs. 9 and 10, respectively. The resistance increased gradually on cooling down from room temperature to a maximum at about 12-13 K, and its maximum value is about half that of the bilayer in Fig. 5 if it were prepared on a whole wafer. Fig. 9 shows a spread of the $T_c$ values for the different contact locations on the wafer, with transition onset at about 6 K. The MR data versus temperature of the 10 nm Au on 4 nm NbN bilayer is shown in Fig. 10. The inset of this figure shows the MR at low temperatures under 2 and 4 T magnetic fields, and under different gate voltages. The main panel is a zoom-in on the tail of the MR of the inset up to 90 K. The increase above about 1\% of the MR in the inset marks $T_c\cong 6$ K of the main part of the bilayer which coincides with the resistive $T_c$ onset value obtained in Fig. 9. Below this $T_c$ value, the MR increases rapidly with decreasing temperature due to flux-flow in the bilayer. Between 4 and 5 K, one can observe a distinct knee in the MR which indicates the transition between superconductivity in the NbN islands above 5 K, and the proximity induced superconductivity in the gold layer in between islands below 4 K. The saturation MR at 200 \% is a result of our definition of MR, which get this value when R(0)=0. These features are similar to those of the inset to Fig. 5 in a 10 nm $Bi_2Se_3$ on 3 nm NbN bilayer under 0 Vg, where a lower $T_c$ of 2.7 K was found, with a weaker signature of the NbN islands and inter-islands regimes, and without the saturation effect since the bilayer always remained resistive, even at 1.8 K. The inset to Fig. 10 also shows a small gating effect on $T_c$ at 2 T, where under 0 Vg $T_c$ is higher by $\sim$0.3 K than at -100 Vg. \\

\begin{figure} \hspace{-20mm}
\includegraphics[height=6cm,width=8cm]{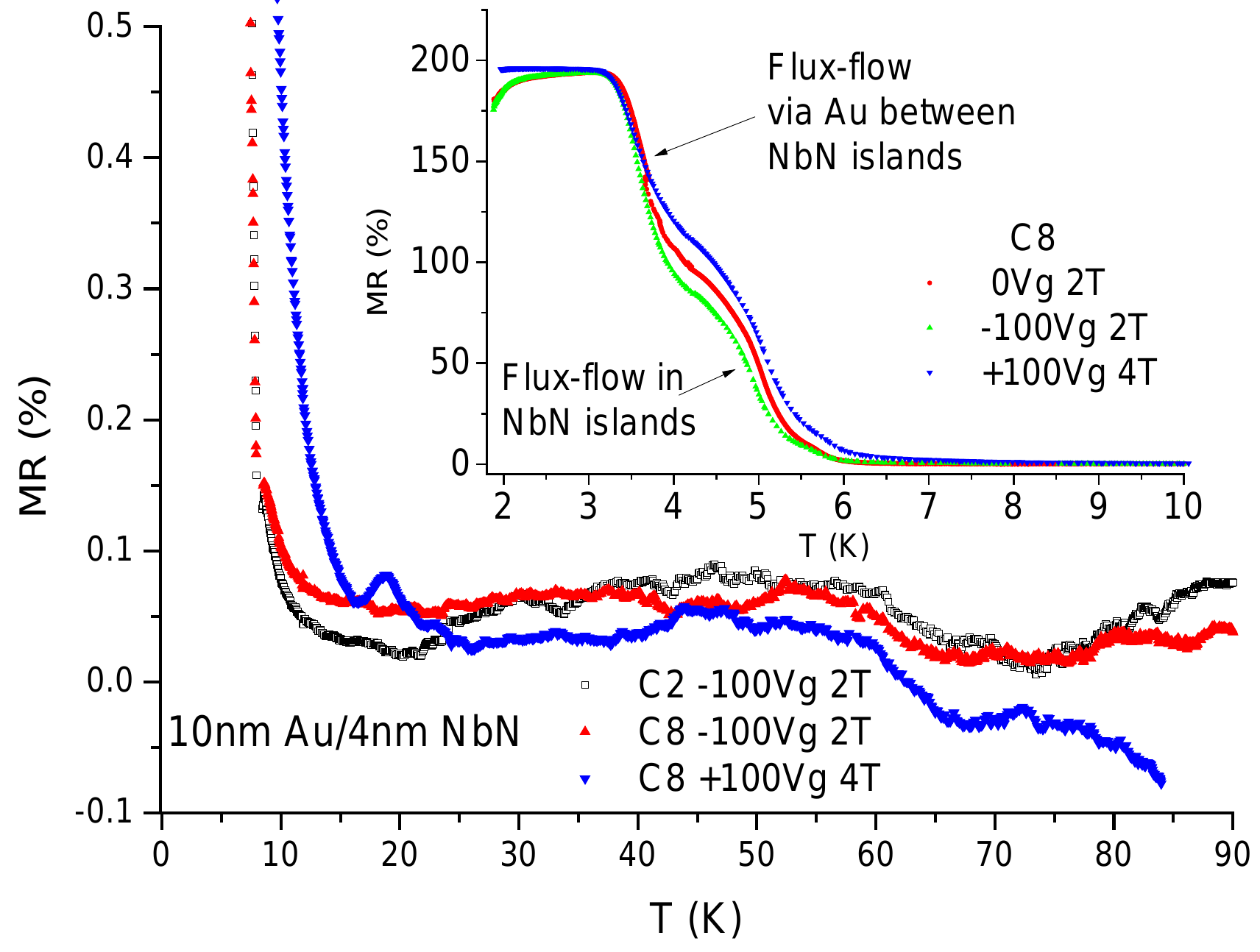}
\hspace{-20mm} \caption{\label{fig:epsart} (Color online) Magnetoresistance versus temperature of the bilayer of Fig. 9 under different magnetic fields and gate voltages.
The inset shows the low temperature behavior of the MR below $T_c\sim 6$ K.}
\end{figure}

The main panel of Fig. 10 shows that the MR above 12-13 K and up to 90 K is within the noise of the measurements ($\pm$0.1 \%). This is in contrast to the observation in the bilayer of Fig. 6 where an MR of a few percents is observed even much above $T_c$. If we  determine $T_c$ in Fig. 10 (somewhat arbitrarily) at T where MR$> 0.2$\%, we find $T_c$ values of 8 and 12 K at 2 and 4 T fields, respectively. These $T_c$ values mark the onset of flux-flow at the given fields, and are higher than the previously determined $T_c$ from the inset ($\sim$6 K). All these different $T_c$ values reflect the fact that MR values higher than the noise level (reached at the onsets of flux-flow) occur at higher temperatures under higher fields, and depend sensitively on the onset criterion.  In addition, the $T_c = 12$ K value coincides with the temperature value of maximum resistance as seen in Fig. 9. Therefore, we conclude that this is the $T_c$ of the thicker NbN islands in the bilayer. \\

\subsection{3.4. $Bi_2Se_3-NbN$ bilayers}

We now present the main part of this study, namely, the properties of a highly resistive 10 nm $Bi_2Se_3$ on 3 nm NbN bilayer, and this after an extensive preparatory part on the properties of films and bilayers pertinent to the present investigation. It turned out that the results of the R and MR in Figs. 5 and 6 of a similar bilayer (on half a wafer) are not very interesting, since the resulting MR  above $T_c$ is very similar to that of a bare 10 nm $Bi_2Se_3$ film. This is a direct result from the fact that the links between superconducting islands of the 3 nm reference NbN film were still too strong, as is obvious from the almost full transition to zero resistance at low temperatures. We therefore had to weaken the inter-grain links even more. Since the use of 2.5 nm thick NbN films on FS yielded very resistive films, with R over a M$\Omega$ at low temperatures, with  a lot of noise and a very weak signature of a transition at 2-3 K, we decided to stick with the 3 nm thick films in the bilayers. We thus prepared and characterized 7 wafers with the 10 nm $Bi_2Se_3$ on 3 nm NbN bilayers, and found out that their maximum resistance (at low temperatures) varied between about 1 and 10 k$\Omega$. This depended mostly on the base pressure used before the deposition process (1.5-3$\times 10^{-7}$ Torr), but probably also on the surface quality of the FS wafers. In any case, the most interesting results were obtained with the wafers of highest resistance, and here we present the results on the most resistive one, which also has the weakest links as we originally planned to have. \\

\begin{figure} \hspace{-20mm}
\includegraphics[height=6cm,width=8cm]{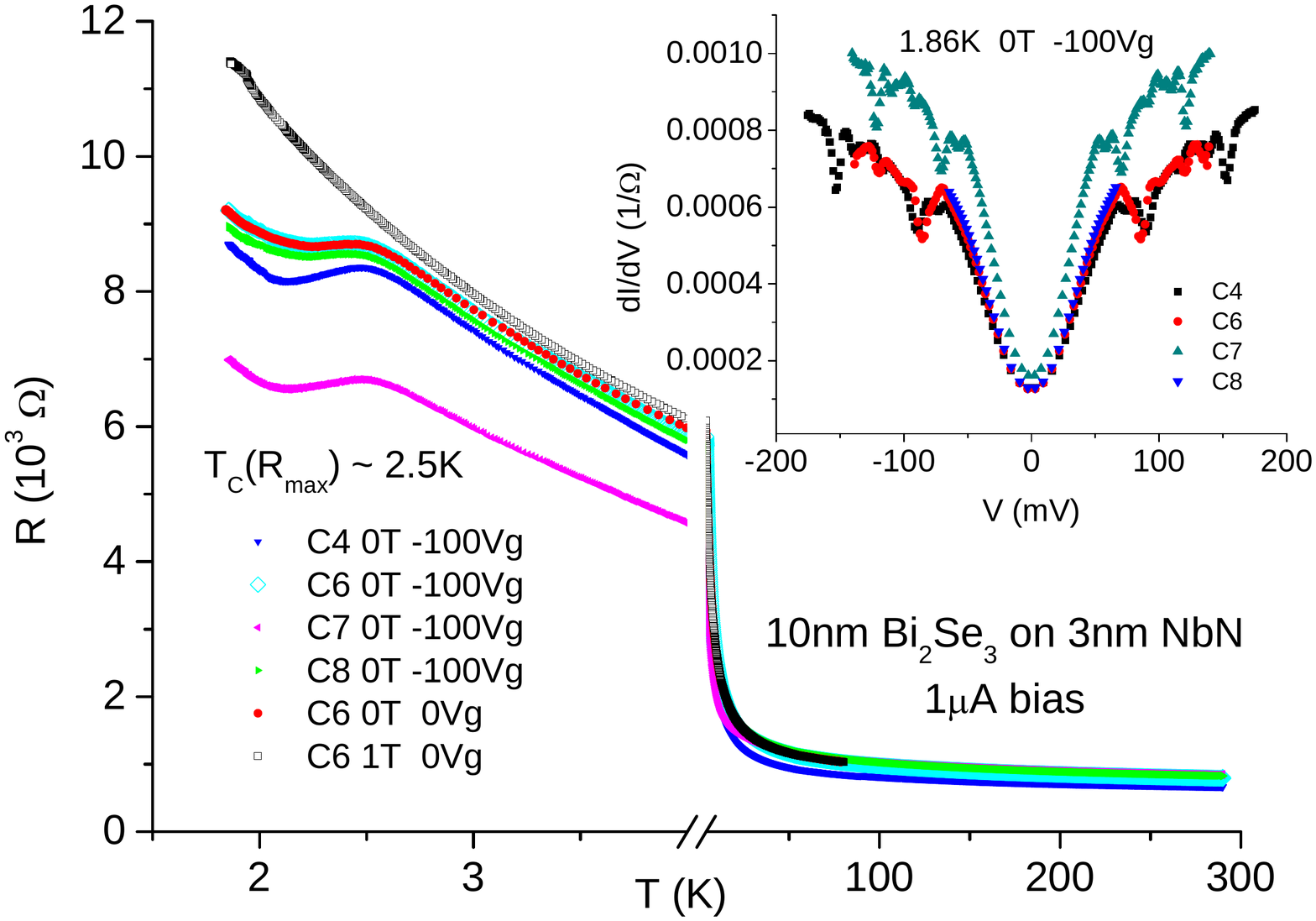}
\hspace{-20mm} \caption{\label{fig:epsart} (Color online) Resistance versus temperature of a pristine 10 nm $Bi_2Se_3$ on 3 nm NbN on FS bilayer under different magnetic fields and gate voltages. The inset shows conductance spectra of this highly resistive bilayer at 1.86 K, 0 T and -100 Vg.  }
\end{figure}

Fig. 11 shows the resistance versus temperature of this wafer with the 10 nm $Bi_2Se_3$ on 3 nm NbN on FS bilayer. The temperature dependence of R clearly shows insulating behavior down to 2.5 K where a small decrease of resistance is observed. Similar behavior was observed also in 5 nm thin cuprate films \cite{KM2}. We show data at four locations on the wafer which exhibits quite a spread, and also the data of contact C6 at 0 and 1 T magnetic field under 0 Vg. Clearly, the small decrease of R below 2.5 K disappears under 1 T, and we thus conclude that it originates in superconductivity of the NbN islands which are very weakly linked. The inset shows conductance spectra of all these contacts at 1.86 K which clearly exhibit tunneling behavior between the superconducting islands. The gap-like feature appearing at about 70 meV, is a result of several weak-links connected in series between the voltage contacts. Assuming an energy gap $\Delta$ of 2 meV for the NbN islands \cite{Kamlapure} yields 35 such weak-links in series.  The dips in the conductance spectra of the inset are due to heating effects as a result of reaching the critical current of the NbN islands \cite{IcDips}. \\

\begin{figure} \hspace{-20mm}
\includegraphics[height=6cm,width=8cm]{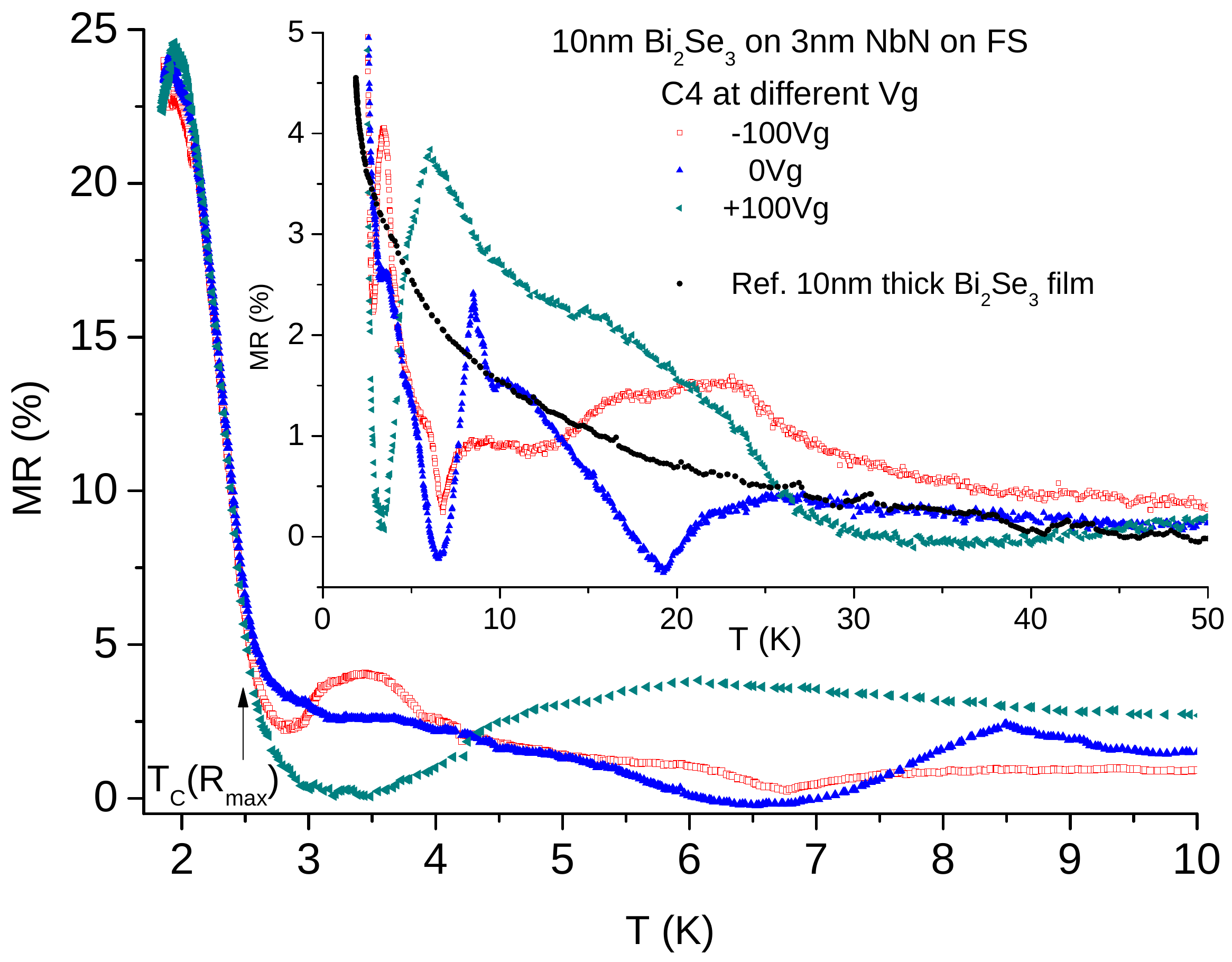}
\hspace{-20mm} \caption{\label{fig:epsart} (Color online) Magnetoresistance versus temperature of the C4 contact of the pristine bilayer of Fig. 11, obtained for 0 and 1 T fields and under gate voltages of 0 and $\pm100$ Vg. All other contacts show similar behavior. The inset shows the full temperature range while the main panel is a zoom in on the low temperature regime. The inset also shows the MR of the reference 10 nm thick $Bi_2Se_3$ film of Fig. 6.
}
\end{figure}

The main results of this study are presented in Fig. 12 and its inset, where the MR of the bilayer of Fig. 11 for 0 and 1 T and under different gate voltages shows a complex, non monotonous peaks structure. For comparison, the corresponding monotonous MR of the reference 10 nm $Bi_2Se_3$ film of Fig. 6 is also shown. Below 2.5 K, the MR is similar to the results in the inset to Fig. 5, but with a 3-5 times smaller signal at the maximum MR at $\sim$1.9 K. The later reflects the fact that the weak-links here are much weaker than in Fig. 5. There is also no visible effect of the gate voltage on the MR results in this regime. Above 2.5 K, the MR data becomes very sensitive to the gate voltage, and up to 7-8 K exhibit kind of anti-phase behavior versus temperature for $\pm$100 Vg where the peak and dip at 3.5 K reverse roles at 6-7 K. The MR under 0 Vg is somewhere in between these two at 3.5 K, but then becomes closer to the -100 Vg data at 6-7 K. At higher temperatures the MR data becomes even more complex, but clearly it goes down at 25-30 K, where it tends to follow the reference $Bi_2Se_3$ film. It should be stressed here that the MR behavior of Fig. 12 for the C4 contact, was observed also for the other contacts on this wafer (C6 and C8), while C7 had an additional strong oscillatory behavior above 10 K as shown in Fig. 13 for 0 Vg. This oscillatory behavior was washed out under $\pm$100 Vg. The inset to Fig. 13 is a zoom-in on weaker oscillations or plateaus in the MR of C7 in the range of 3-8 K. Such plateaus and knees, although weaker, appear also in the other contacts. As far as we know, there is no theory that predicts oscillations of MR versus temperature in a 2D topological superconductor.  \\

\begin{figure} \hspace{-20mm}
\includegraphics[height=6cm,width=8cm]{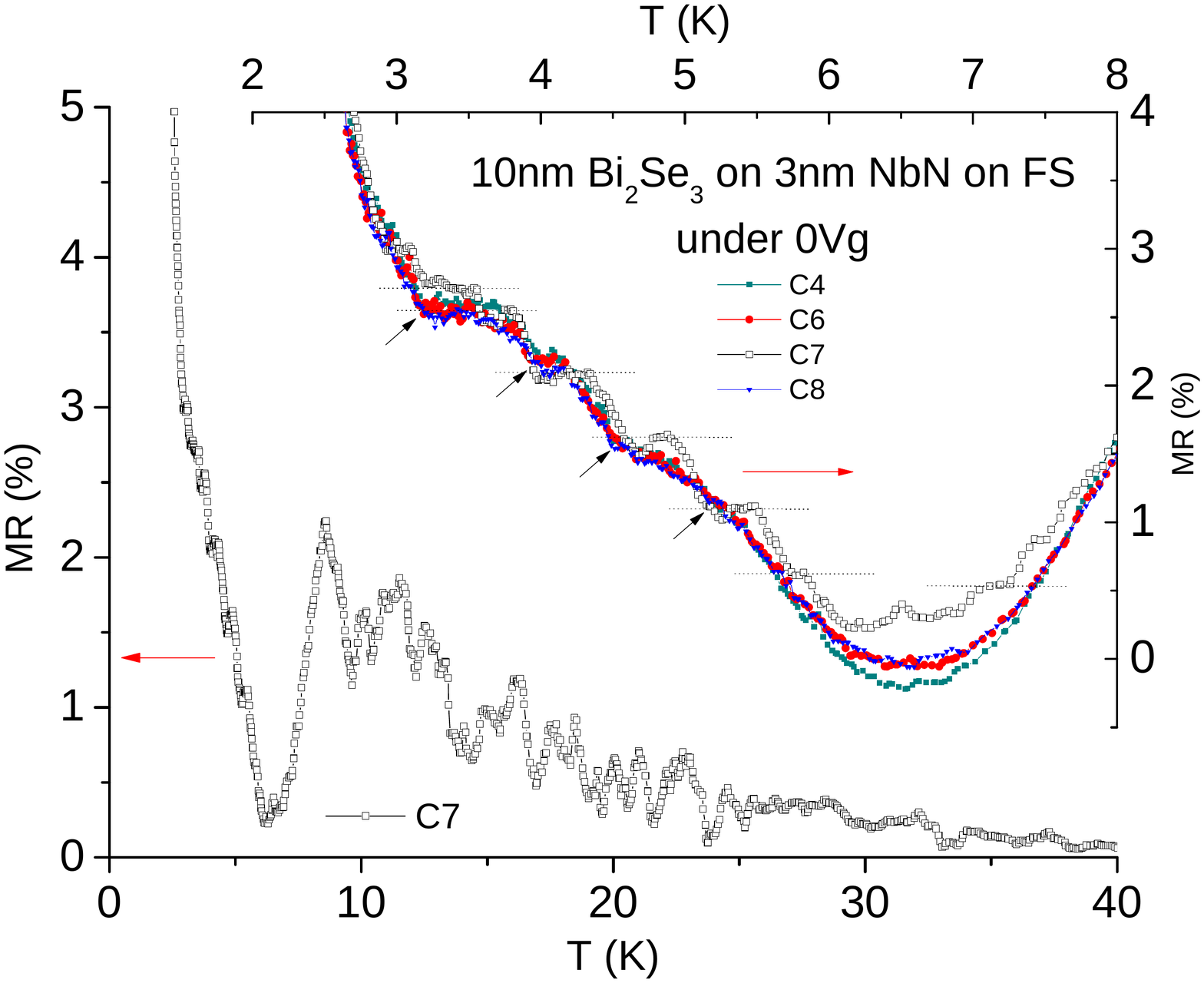}
\hspace{-20mm} \caption{\label{fig:epsart} (Color online) MR versus temperature at 0Vg of the C7 contact on the wafer of Fig. 11. The oscillations were smeared out under $\pm 100$ Vg. The inset is a zoom in on the data at low temperatures. Only this C7 contact showed the pronounced oscillations versus temperature, although plateaus and less pronounced knees were observed also in the other contacts as seen in the inset and marked by the arrows.  }
\end{figure}

\begin{figure} \hspace{-20mm}
\includegraphics[height=6cm,width=8cm]{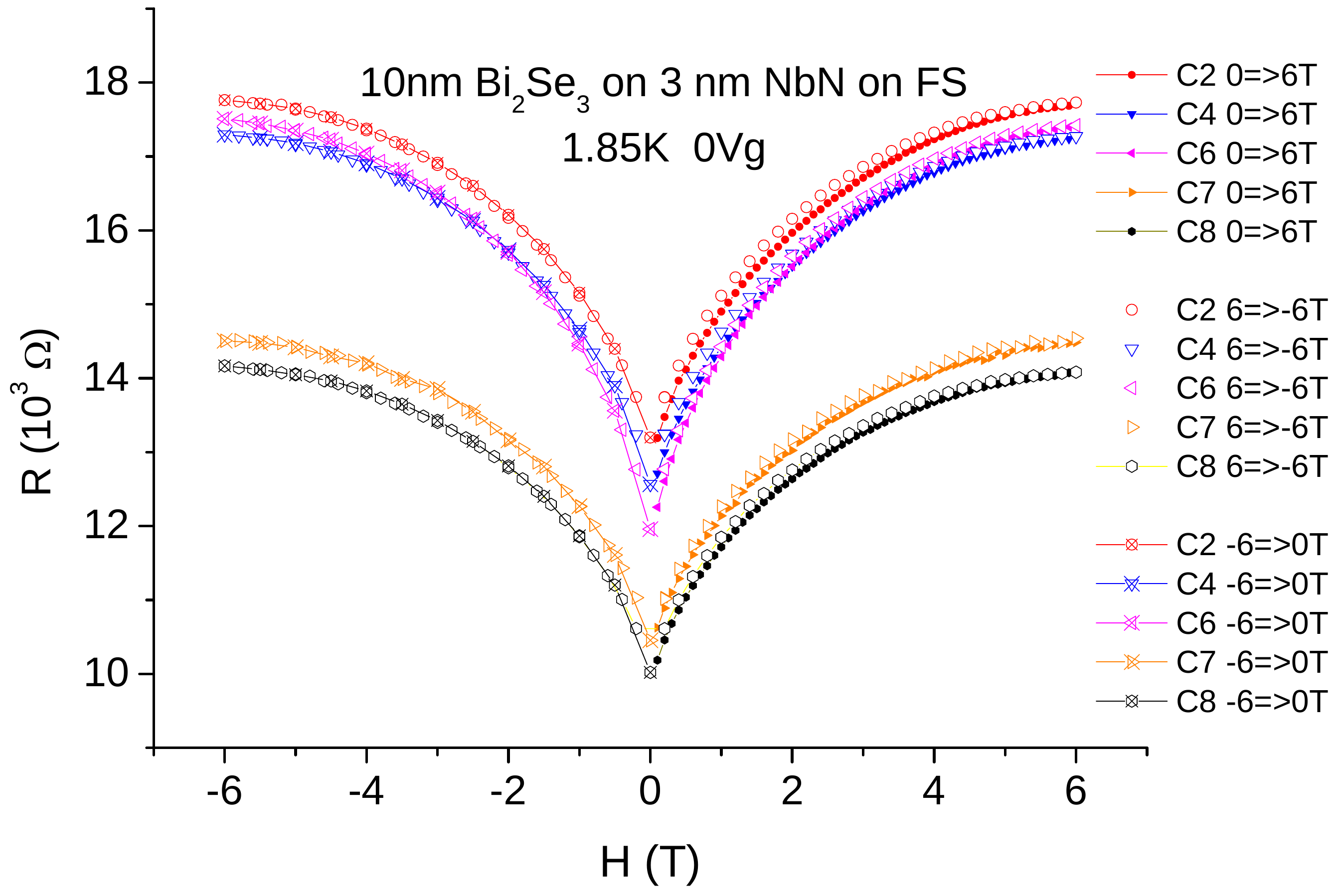}
\hspace{-20mm} \caption{\label{fig:epsart} (Color online) Resistance versus magnetic field of the bilayer of Fig. 11 at 1.85 K and 0 Vg. Although this bilayer was kept in dry air for one month prior to this measurement, it deteriorated (aged) as can be seen from the increase of its resistance values at 0 T by about 4 k$\Omega$ as compared to Fig. 11.  }
\end{figure}

\begin{figure} \hspace{-20mm}
\includegraphics[height=6cm,width=8cm]{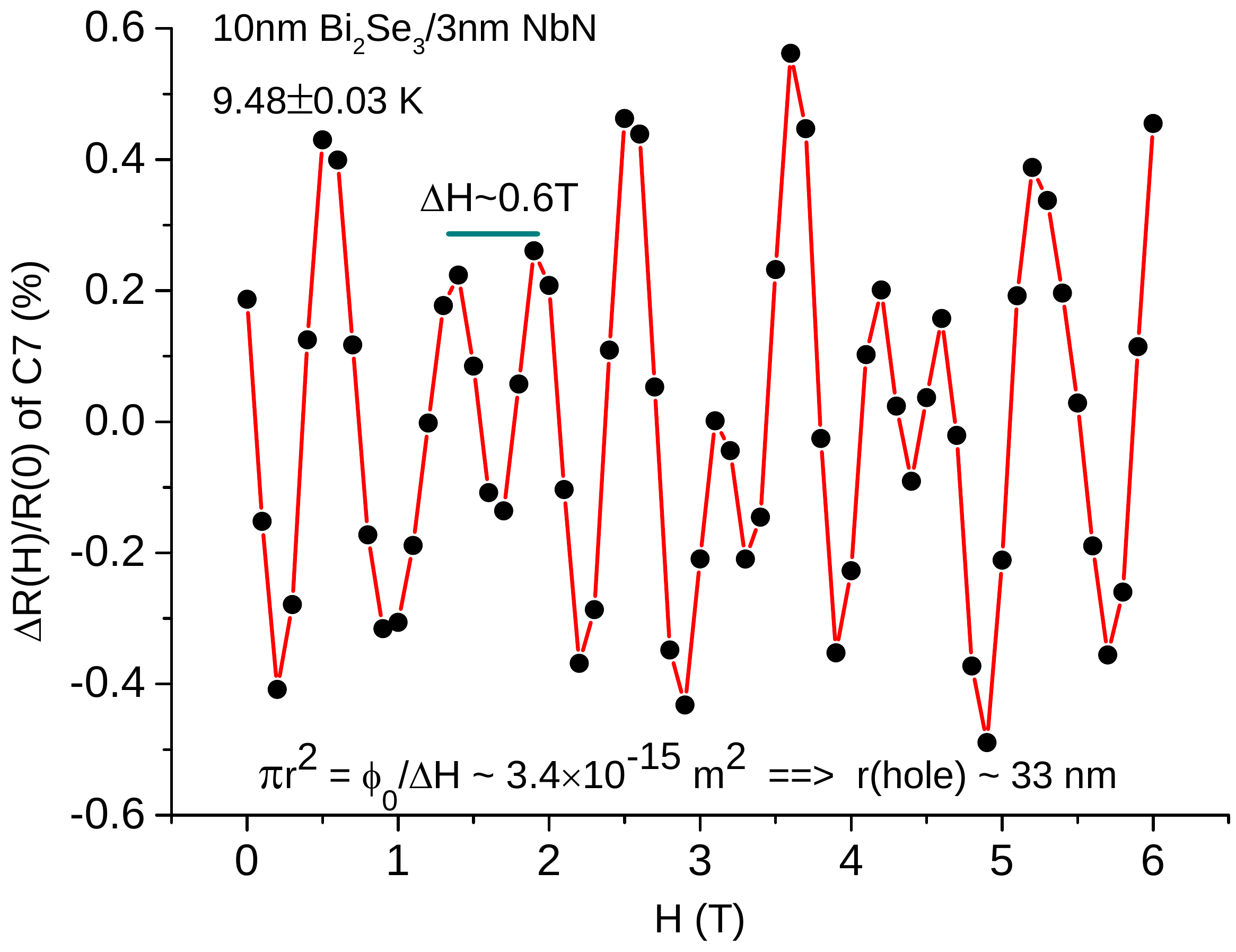}
\hspace{-20mm} \caption{\label{fig:epsart} (Color online) Oscillations of normalized R minus background [$\Delta$R(H)/R(0)] of the C7 contact of Fig. 13 versus field at 9.48 K and 0 Vg. }
\end{figure}

To find out whether these oscillations and plateaus originate  in magnetic field effects, we measured R versus H for all the contacts on the wafer of Fig. 11 at 1.85 K. The results are depicted in Fig. 14, and they are quite similar to those of the inset to Fig. 6, besides the difference in overall resistance and aging effects. We have also measured R versus H under 0 Vg  on this  wafer (when it was 1.5 month old) at higher temperatures. A background of a polynomial of order 2 fit to the data was subtracted from the data, and the resulting resistance variation $\Delta R$ versus H was obtained. The normalized $\Delta$R(H)/R(0) results at $\sim$9.5 K are plotted in Fig. 15 versus magnetic field. Presuming that the observed oscillations originate in flux quantization in nano-holes in the bilayer, we obtain from the period $\Delta H \sim 0.6$ T a hole diameter of 66 nm, which agrees quite well with the AFM image and line profile of Fig. 1. This result suggests that superconductivity in the bilayer persists at least up to $\sim$9.5 K. Similar oscillations were observed also in the other contacts at 3.7 and 5.3 K, but they disappeared at 15.1 K in the noise level of the measurements. Thus $T_c$ of the NbN islands in this bilayer should be between 9.5 and 15.1 K. This is in agreement with the the MR onset in Fig. 10 under 4 T (12.2 K), or the temperatures of maximum resistance of Fig. 9 (12-13 K), although in that case the NbN film was 4 nm thick and not 3 nm thick as in the present bilayer. We conclude from these findings that $T_c$ of the superconducting islands in our bilayer can reach 12-13 K, and that if fluctuations are taken into account, superconductivity could persist in our bilayers up to $\sim$15 K. Thus, it should be possible to observe flux-flow effects of vortices  up to this temperature, but this however, does not explain the non monotonous MR features observed in Figs. 12 and 13. Above 15 K, any MR effect different from that due to the reference 10 nm $Bi_2Se_3$ film, should have a different origin. \\

Next, we discuss possible explanations for the complex behavior of the MR results of Fig. 12. In principle, the present bilayer as drawn schematically in Fig. 4, comprises of six distinct superconducting regions, marked in this figure by I to VI. The NbN islands (I) and the inter-islands regions (II) are the obvious ones, but proximity induced superconductivity occurs also in the $Bi_2Se_3$ layer above these regions, in particular, at the  bottom and top surfaces of this doped topological layer where the helical edge currents flow. This yields four more superconductive regions in the cap $Bi_2Se_3$ layer, two above the islands (III and V) and two above the inter-islands regions (IV and VI). Due to proximity effect and distances from the superconducting NbN islands (I), we can safely assume that $T_c$(I)$>T_c$(III)$>T_c$(II)$>T_c$(IV). We shall assume first that the MR results observed in Fig. 12 originates in vortex physics, and discuss them in terms of pinning in the different superconducting regimes I-VI. Then, for the remaining unidentified features, we shall discuss possible contributions to the MR from zero energy Majorana modes and unconventional superconductivity. \\

In the context of vortex physics, the MR peaks seen in Fig. 12 as a function of temperature originate in flux-flow, and are due to two competing phenomena. One is the increased vortex generation on cooling down due to newly formed proximity induced superconducting areas in the bilayer which eventually saturates, and the other is increased pinning with decreasing temperature. The result of these effects is an MR peak versus temperature below which all vortex motion stops and the MR goes to zero. Such MR peaks have been clearly observed in similar cuprate bilayers \cite{KM1}, as well as in the inset to Fig. 5 here. Since the bottom $Bi_2Se_3$ surface of regions III and IV are very close to regions I and II, the proximity induced superconductivity in them is stronger than in the more distant regions V and VI. We shall first discuss the MR versus T result of Fig. 12 under 0 Vg. The MR of this data is most always lower than that of the reference $Bi_2Se_3$ film. This looks like MR suppression in the bilayer, but since there are zero MR regimes in the bilayer data (at 6-7  and 18-20 K), it looks more like an indication of strongly interacting layers in the bilayer. This interaction seems to play a lesser role only above 25 K where the two curves coincide, apparently because no superconductivity affects the MR results at this high temperature range. We attribute the double MR peak between 7-15 K to flux flow, vortex generation and pinning in regions I and III. We note here that the above analysis was facilitated by the fact that this double peak is fully separated from the other features of the MR curve. Using similar arguments, we can attribute the broad MR knee between 2.5 and 6.5 K to pinning in regions II, and the large MR peak below 2.5 K to pinning effects in region IV. The fact that the MR peak below 2.5 K is so strong might be due to enhance proximity induced superconductivity by the surface currents of the topological layer in region IV, or to additional contributions from other effects, to be discussed below. \\

We shall now analyze the MR results of Fig. 12 under different gate voltages and try to explain the two prominent peaks at 3.5 and 6 K under -100 Vg and +100 Vg, respectively. We note that transport, and therefore also the MR data, is strongly dependent on the conductance of regions II and IV, since they comprise the weak-links for current flow in the bilayer. As a result, they will be more sensitive to gating than region I.  Under -100 Vg,  the depletion layer in Fig. 4 in between the NbN islands (region II), is positively charged or simply electron depleted. This makes region II, (as well as the adjacent region IV just above it), less conducting (or more insulating), thus lowering its $T_{c}$ onset to about 4 K compared to its value without gating (onset at $\sim$6.5 K). The MR peak at 3.5 K under -100 Vg is consistent with this scenario, and therefore seems to originate in region II. The large MR peak at 1.9 K seems to be unaffected by either of the present gate voltages of $\pm$100 Vg, and this again might originate in topological effects in region IV. Under +100 Vg, the depletion layer is electron rich (more conducting), thus its proximity induced $T_{c}$ in region II is higher than the 4 K obtain before under -100 Vg. The MR peak at 6 K under +100 Vg can thus originate also in region II. Fig. 12 shows extra MR features under gating besides the two prominent peaks at 3.5 and 6 K. These include the broad hump under +100 Vg between 10 and 25 K, and the broad peak between 15 and 30 K under -100 Vg. As discussed before, superconductivity dies off at about 15 K, and any proximity induced $T_c$ will obviously be lower. Therefore, if one wishes to invoke vortex motion as the origin of these extra MR features, one has to assume the existence of a pseudogap phase for the present bilayers. Signature of such a phase was found in point contact conductance spectra measurements of copper doped $Cu_{0.2}Bi_2Se_3$ single crystals with $T_c\sim$3 K, where a depletion of density of states at low bias was observed up to $\sim$20 K \cite{Kirzhner}.
Considering that in the present bilayers the NbN islands have a much higher $T_c\sim$12-13 K, the conjectured pseudogap phase could easily reach 25-30 K. Hence, to summarize, the extra MR features in Fig. 12 above 15 K could originate in flux flow, provided a pseudogap phase exists in our system. In the cuprates, where the pseudogap phase is well established, such vortices were detected in thermoelectric measurements under a magnetic field (the Nernst effect) \cite{Ong}. \\

Finally, we focus on the intensity or magnitude of the MR peaks in Fig. 12, and on what might contribute extra strength to the dominant peak at 1.9 K. We note that the volume fraction of the superconducting NbN islands with the highest $T_c$ values is apparently very small, as can be inferred from the very small resistive transition at 2.5 K in Fig. 11. Previously, in the vortex pinning context, we attributed the MR double peak under 0 Vg at 7-15 K to regions I and III, the knee between 2.5 and 6.5 K to region II, and the large MR peak below 2.5 K to region IV. It was hard to reconcile how region IV, with its presumably weakest superconductivity, leads to such a large MR peak. Thus, if the enhanced MR peak at 1.9 K is still due to pinning effects, we are led to the conclusion that region IV must have enhanced superconductivity just as well. Such enhancement might originate in a longer normal coherence length $\xi_N$ in the $Bi_2Se_3$ at low temperatures, which is possibly enhanced further by hybridization of the helical surface currents of regions IV and VI. We know of no theory that predicts such effects, and therefore this interpretation is only a hypothesis at the present time. An alternative scenario for the interpretation of our results is that the MR knee between 2.5 and 6.5 K is due to pinning effects in both region II and region IV. This would leave the dominant MR peak at 1.9 K unaccounted for, and other effects as for its origin should be explored. But this time,  vortices in the bilayer are fully pinned and presumably form a vortex lattice. Vortices in general, in a topological superconductor, are predicted to host zero energy Majorana modes in the vortex cores \cite{FuKane}. Thus, in a dense periodic vortex lattice, interactions between these modes lead to the creation of new energy bands in the band gap of this system, some of which could be quite flat \cite{Franz}.  Whether or not these Majorana bands contribute to the MR of the bilayer is unknown at the present time. Rigorous MR calculation versus temperature are needed in order to test if this scenario actually occurs in the present bilayers. \\

\section{4. Conclusions}

A comprehensive study of ultra thin bilayers comprising of weakly connected s-wave superconducting islands and a continuous topological insulator cap-layer reveal interesting gate and temperature dependent magnetoresistance features. These features could be explained as originating in vortex physics, where proximity induced superconductivity and pinning effects in the different regions of the bilayer play a major role. In particular, the present results are consistent with enhanced superconductivity in the surface region of the topological layer in contact with the superconductor, where the helical edge currents flow. Signature of a pseudogap was found, and a possible contribution to the data from Majorana bands was discussed.\\

{\em Acknowledgments:}  We greatly acknowledge useful discussions with Felix von Oppen and Netanel Lindner.\\

\bibliography{AndDepBib.bib}

\bibliography{apssamp}

\begin{thebibliography}{99}
\label{Bib}

\bibitem{KaneRMP} M. Z. Hassan and C. L. Kane,  \textit{Rev. Mod. Phys.}  \textbf{82} 3045 (2010).

\bibitem{FuKane} Liang Fu and C. L. Kane \textit{Phys. Rev. Lett.} \textbf{100} 096407 (2008).

\bibitem{Pesin} D. Pesin and A. H. MacDonald,  \textit{Nat. Mattter.} \textbf{11} 409 (2012).

\bibitem{Kitaev} A. Yu Kitaev,  \textit{Annals of Physics} \textbf{303} 2 (2003).

\bibitem{Kriener} M. Kriener, K. Segawa, Ren Zhi, S. Sasaki, S. Wada, S. Kuwabata and Y. Ando, \textit{Phys. Rev.  B} \textbf{84} 054513 (2011).

\bibitem{AndoRev} Y. Ando,  \textit{J. Phys. Soc. Jpn.} \textbf{82} 102001 (2013).

\bibitem{Koren1} G. Koren, T. Kirzhner, E. Lahoud, K. B. Chashka and A. Kanigel,  \textit{Phys. Rev. B} \textbf{84} 224521 (2011).

\bibitem{LiLu} Yang Fan, Y. Ding, F. Qu, J. Shen, Jun Chen, Z. Wei,
Z. Ji, G. Liu, J. Fan, C. Yang, T. Xiang and Li Lu,  \textit{Phys. Rev. B} \textbf{85} 104508 (2012).

\bibitem{Kouwenhoven} V. Mourik, K. Zuo, S. M. Frolov, S. R. Plissard, E. P. A. M. Bakkers  and L. P. Kouwenhoven,  \textit{Science} \textbf{336} 1003 (2012).

\bibitem{Heibloom} A. Das, Y. Ronen, Y. Most, Y. Oreg, M. Heiblum and H. Shtrikman,
\textit{Nature Phys.} \textbf{8} 887 (2012).

\bibitem{Xu} Xu Jin-Peng \textit{et al.}, Phys. Rev. Lett. \textbf{114}, 017001 (2015).

\bibitem{Berg} A. Haim, E. Berg, F. von Oppen and Y. Oreg, Phys. Rev. Lett. \textbf{114}, 166406 (2015).

\bibitem{Brouwer} G. Kells, D. Meidan and P. W. Brouwer,  \textit{Phys. Rev. B} \textbf{86} 100503(R) (2012).

\bibitem{Marcus} A. P. Higginbotham \textit{et al.}, arXiv:1501.05155 [cond-mat.mes-hall] (2015).

\bibitem{Churchill} H. O. H. Churchill, V. Fatemi, K. Grove-Rasmussen, M. T. Deng, P. Caroff, H. Q. Xu and C. M. Marcus,  \textit{Phys. Rev. B} \textbf{87} 241401(R) (2013).

\bibitem{KorenSUST} Gad Koren,  Supercond. Sci. Technol. \textbf{28} 025003 (2015), arXiv:1409.2975.

\bibitem{Koren2} G. Koren and T. Kirzhner,  \textit{Phys. Rev. B} \textbf{86} 144508 (2012).

\bibitem{Koren3} G. Koren, T. Kirzhner, Y. Kalcheim and O. Millo,  \textit{EPL} \textbf{103} 67010 (2013).

\bibitem{Xin-kang} D. Xin-kang, W. Tian-min, W. Cong, C. Bu-liang, Z. Long,  \textit{Chinese. J. Aero.} \textbf{20} 140 (2007).

\bibitem{Darlinski} A. Darlinski and J. Halbritter,  \textit{Surf. Interface Anal.} \textbf{10} 223 (1987).

\bibitem{Butch} N. P. Butch, K. Kirshenbaum, P. Syers, A. B. Sushkov, G. S. Jenkins, H. D. Drew and J. Paglione,  \textit{Phys. Rev. B} \textbf{81} 241301(R) (2010).

\bibitem{Steinberg} H. Steinberg, J. B. Laloe, V. Fatemi, J. S. Moodera, and P. Jarillo-Herrero, Phys. Rev. B \textbf{84}, 233101 (2011).

\bibitem{Bergmann} G. Bergmann,  Solid State Commun. \textbf{42}, 815 (1982); Phys. Rep. \textbf{107}, 1 (1984).

\bibitem{HLN} S. Hikami, A. I. Larkin and Y. Nagaoka, Prog. Theor. Phys. \textbf{63}, 707 (1980).

\bibitem{KM2} G. Koren and O. Millo,  \textit{Phys. Rev. B} \textbf{80} 054507 (2009).

\bibitem{Kamlapure} A. Kamlapure \textit{et al}., Appl. Phys. Lett. \textbf{96}, 072509 (2010).

\bibitem{IcDips} G. Sheet, S. Mukhopadhyay and P. Raychaudhuri,  \textit{Phys. Rev. B} \textbf{69} 134507 (2004).

\bibitem{KM1} G. Koren and O. Millo, Phys. Rev. B \textbf{81}, 134516 (2010).

\bibitem{Kirzhner} T. Kirzhner, E. Lahoud, K. B. Chaska, Z. Salman, and A. Kanigel, Phys. Rev. B \textbf{86}, 064517 (2012).

\bibitem{Ong} Yayu Wang, Lu Li and N. P. Ong, Phys. Rev. B \textbf{73}, 024510 (2006).

\bibitem{Franz} Tianyu Liu and M. Franz, arXiv:1506.05084v1, (2015).



\end{thebibliography}

\end{document}